\documentclass[runningheads]{llncs}
\usepackage{graphicx}
\usepackage{xcolor}
\usepackage{amsfonts}
\usepackage{hyperref}
\usepackage{booktabs}
\usepackage{enumitem}
\usepackage{amsmath}
\usepackage{subfig}
\usepackage{floatrow}
\newfloatcommand{capbtabbox}{table}[][\FBwidth]

\begin{document}
\title{Trenchcoat: Human-Computable Hashing Algorithms for Password Generation}
\titlerunning{Trenchcoat: Human-Hashing for Password Generation}
%
\author{Ruthu Hulikal Rooparaghunath\orcidID{0000-0002-7770-4028}
\and T.S. Harikrishnan
\and Debayan Gupta\orcidID{0000-0002-4457-1556}}
\authorrunning{R. Rooparaghunath et al.}
%
\institute{
    Ashoka University, Haryana, India,\\
    \email{ruthu.rooparaghunath\_ug21@ashoka.edu.in}, 
    \email{ts.harikrishnan@alumni.ashoka.edu.in}, 
    \email{debayan.gupta@ashoka.edu.in}
}
\maketitle              

\newcommand{\debayan}[1]{\textcolor{red}{#1}}
\newcommand{\ruthusnotetoself}[1]{\textcolor{blue}{#1}}

\begin{abstract}

The average user has between 90-130 online accounts~\cite{digital}, and around $3\times10^{11}$ passwords are in use this year~\cite{300billionpasswords}. Most people are terrible at remembering ``random'' passwords, so they reuse or create similar passwords using a combination of predictable words, numbers, and symbols~\cite{harrispoll}. Previous password-generation or management protocols have imposed so large a cognitive load that users have abandoned them in favor of insecure yet simpler methods (\textit{e.g.}, writing them down or reusing minor variants). 

We describe a range of candidate \textit{human-computable} ``hash" functions suitable for use as password generators - as long as the human (with minimal education assumptions) keeps a single, easily-memorizable `master' secret - and rate them by various metrics, including \textit{effective security}. 

These functions hash master-secrets with user accounts to produce sub-secrets that can be used as passwords; $F_R($s$, w) \longrightarrow y$, which takes a website $w$ and produces a password $y$, parameterized by the master secret $s$, \textit{which may or may not be a string}.

We exploit the unique configuration $R$ of each user's associative and implicit memory (detailed in section~\ref{sec:cogsci}) to ensure that sources of randomness unique to each user are present in each $F$. An adversary cannot compute or verify $F_R$ efficiently since $R$ is unique to each individual; in that sense, our hash function is similar to a physically unclonable function~\cite{suh2007physical}.  For the algorithms we propose, the user need only complete primitive operations such as addition, spatial navigation or searching.  \textit{Critically, most of our methods are also accessible to neurodiverse, or cognitively or physically differently-abled persons.}

Given the nature of these functions, it is not possible to directly use traditional cryptographic methods for analysis; so, we use an array of approaches, mainly related to entropy, to illustrate and analyze the same.

We draw on cognitive, neuroscientific, and cryptographic research to use these functions as improved password management and creation systems, and present results from a survey (n=134 individuals, with each candidate performing 2 schemes) investigating real-world usage of these methods and how people \textit{currently} come up with their passwords. We also survey 400 websites to collate current password advice.

\end{abstract}


\section{Introduction}

\textit{Your password must be between 8-16 characters long, with at least one uppercase character, one lowercase character, one number, and one special character (such as !,@,\#,etc.), must not include your username, and be changed every 90 days.}

\medskip

Memorizing myriad passwords, with (often questionable) constraints imposed to make each password as ``random'' as possible, and little guidance on how to manage this information, is a herculean task. This has resulted in people using easily guessable and common passwords~\cite{SplashData}. Surveys last year indicated that individuals reuse over half of all passwords for multiple accounts, with many others being easily attacked with a dictionary of common passwords~\cite{harrispoll}.

Anecdotally, users prioritize convenience over privacy when accessing newsletters, spam mails, or magazine subscriptions. They assign important accounts with less conveniently memorable passwords. This trade-off in memorability results in compromised security when passwords are written and stored at home.~\cite{smith_2020} Weak passwords are a serious threat when they guard sensitive data or systems, and may lead to identity theft, insurance fraud, public humiliation, etc. ~\cite{zetter_2017}.

Common approaches to handling this rely on instructing users to create `strong' passwords with suggestions such as: `don't use your name or birth-date', `include symbols' and `don't capitalize only the first letter'. However, users routinely ignore or circumvent these suggestions because of their cognitive load.

The current standard for password management and security is a password manager. Unfortunately, several sources report serious flaws (including zero-day attacks) \textit{consistently} found in the most popular password managers every year~\cite{gedeon_2020,bestreviews_2018}. Some managers are also vulnerable because of their tendency to store the passwords to the password manager in plaintext.\footnote{Preventing this, in most password managers, requires users to terminate the manager each time after use. Users may be unaware of this or disregard it because of inconvenience, which once again lowers its security~\cite{o'flaherty_2019}.} 

Digital and physical copies of passwords will always have vulnerabilities, but remembering several passwords imposes a cognitive load that users are unwilling or unable to manage. Past research has proposed several password generation methods~\cite{CuePinSelect,blocki2013naturally,blocki2014towards} but those that consider real-world usage have not been tested beyond a dozen people~\cite{CuePinSelect}, or have placed too large a cognitive load on users.

We propose a family of \textbf{public} derivation functions $F$ such that, if we start with a master secret, $s$ (which the human memorizes), we can derive a sub-secret $y_i$ for each website $w_i$. Broadly, our requirements for such $F$ would be: (1) Given $(y_i, w_i)$, where $y_i= F_R(s, w_i)$, it should be computationally hard to find $s$; (2) Given $(y_1, w_1), (y_2, w_2), \dots, (y_k, w_k)$ and $w_{k+1}$, it should be computationally hard to find $y_{k+1}$ (secure as in Unforgeability under Random Challenge Attack~\cite{blocki2014towards}). This minimizes cognitive load by requiring only the memorization of $s$, with any $y_i$ being derived using public $w_i$ and $F$.~\textit{Critically, $s$, unlike $y_i$, need not be a string}! (We discuss visual and cue-based $s$ in section~\ref{sec:cogsci}.)

$F$ must be easily human-computable. F must also not require too much aid, to minimize cognitive load. Further, for individuals to reproduce the same password each time, $F$ should be deterministic with respect to each individual. One way to satisfy most of these requirements is through a cryptographically secure hash.

Predefined cryptographic hash functions such as SHA-3 (with preset size parameters, and conversion to appropriate characters) could be used in place of $F$, calculating $y = F(s \cdot w)$, concatenating $s$ and $w$ where $s$ is a string.
Unfortunately, most humans cannot easily compute SHA-3 in their heads. We need something that includes \textit{some} features of a cryptographically-secure hash function without requiring the mathematical heavy-lifting common to such schemes. In the rest of this document, we describe a number of approaches to finding the same, and the results of our survey on the subject. (Assumptions made by cryptographers on what laypersons would find ``easy to compute'' may be  incorrect; we must empirically observe the methods people are willing and able to use.)

\subsection{Paper Outline and Contributions}

To optimize our hash functions for human use, we discuss visual cues, and implicit and associative methods suggested by cognitive and neuro-scientific research in section~\ref{sec:cogsci}. Previous literature on human-computable passwords requires rehearsal schedules, online aid, etc. with various caveats and problems~\cite{CuePinSelect,blocki2013naturally,blocki2014towards}. 

These issues are obviated by using an easily-memorized key with human-computable algorithms designed for password generation and management. Section~\ref{sec:function-descriptions} presents a range of such hashing algorithms. An adversary cannot compute or verify these hashes efficiently, since these are unique to each individual; in that sense, our hash function is similar to a physically unclonable function~\cite{suh2007physical}

In this context, we discuss \textit{effective security} in section \ref{sec:analyse_effective} which weighs cryptographically evaluated security against human usability. \textit{E.g.}, generating random passwords without associative memory techniques or computational tools and writing materials may impose large cognitive loads, reducing usability\footnote{In general, as a human-computable hash function grows in difficulty, a human is more likely to abandon it~\cite{harrispoll,SplashData} and revert to weak password practices. So, one can have very high theoretical security but, in practice, be totally insecure.}

We also define~\textit{graceful degradation} - our algorithms retain a significant amount of their effective security even if access to writing materials, computers, or the internet is unavailable. We test the algorithms presented in this paper as well as Cue Pin Select~\cite{CuePinSelect} on a survey population of 134 individuals (with each person assigned to two, randomly-chosen schemes), averaging 56 responses per algorithm from people between the ages of 18 to 25. We analyse the results in section~\ref{sec:generation_retention} and also use an LSTM to test character predictability in section~\ref{sec:ml_analysis}.

We cannot use standard cryptographic techniques to evaluate our schemes, as they are explicitly optimized for representation in human brains but difficult to represent or simulate on computers (thus contributing to their security). So, we introduce metrics to assess the security of human-computable schemes, measure ease of use, rememberability, unforgeability under Random Challenge Attack~\cite{blocki2014towards}, and more, in appendix \ref{sec:appendix}.
We also classify algorithms based on their paradigms, limiting factors, and success of password recall in section \ref{sec:analyse_effective}.

Section~\ref{sec:advice} discusses common password hygiene errors and current password advice; we survey 400 websites and applications for such advice (table~\ref{table:advice}). We also provide insight into real-world methods individuals \textit{currently} use to come up with passwords in section~\ref{sec:user_methods}. Finally, section~\ref{sec:hash-analysis} uses our survey results to understand the determinism or stability of our schemes during real-world usage.

\section{Cognitive and Neuro-scientific  Perspectives}\label{sec:cogsci}

During WW1, before the advent of powerful computers, soldiers used ``trench codes'' to communicate across trenches. These had to designed to be computable by soldiers under pressure without assuming high education levels -- this involved coming up with clever codebooks/manuals\footnote{Beyond careful design, these also included side-channel defenses e.g., the paper material was designed to degrade within a few weeks, ensuring that obsolete codes would not be used, and ``lost'' manuals would lose value quickly.}. Such trench codes had their own problems, of course, but these issues were obviated by the time WW2 came around; ever since then, we have optimized our cryptographic functions (encryption, hashing, etc.) for increasingly-powerful computers, not humans. To design human-computable functions while maintaining security, we must first discuss how to optimize functions for the human brain.

Broadly, the brain manages memory in two categories~\cite{blocki2013naturally}: persistent (e.g., notepads) and associative (human memory). The latter is clearly more secure for password storage and recollection, as elaborated in the Introduction. Password recollection depends on the conscious retrieval of detailed memory, which imposes a large cognitive load (so users create workarounds to ease this load). Relying on visual, implicit and associative memory can ease this cognitive load.

Visual memory is capable of long-term storage of large amounts of detailed information. Implicit, associative memory aids in lasting rapid recall.
However memorizing large amounts of new visual information requires constant rehearsal to become embedded in memory, which is tedious. Fortunately, humans already accumulate a vast amount of long-term information throughout their lives. Subconscious rehearsal repeated over time does \textit{not} feel tedious: drawing on implicit memory - such as repeatedly navigating a house - requires less effort.

Visually cued recollection is easier than explicit recollection~\cite{baddeley1997human}. This is also a more accessible method, as neurologically damaged or disabled patients can succeed at implicit memory tasks, even when they cannot succeed on explicit memory tasks~\cite{schacter1993implicit}. We thus contend that password retention relying on implicit memory retrieval has the potential to be \textit{stable, long-lasting, and equitable}.

Some functions proposed in section ~\ref{sec:function-descriptions} are based on this capacity for detailed storage and fast retrieval in visual memory. The \textit{Memory Palace} method uses visually-cued subkey recollection. This can be further improved by using physical copies of partial visual images for cues, eliminating the cognitive load of remembering visual cues themselves. (See section~\ref{sec:hashing} for details of these protocols.) 

\begin{figure}[ht]
\centering
\includegraphics[width= 0.8 \textwidth]{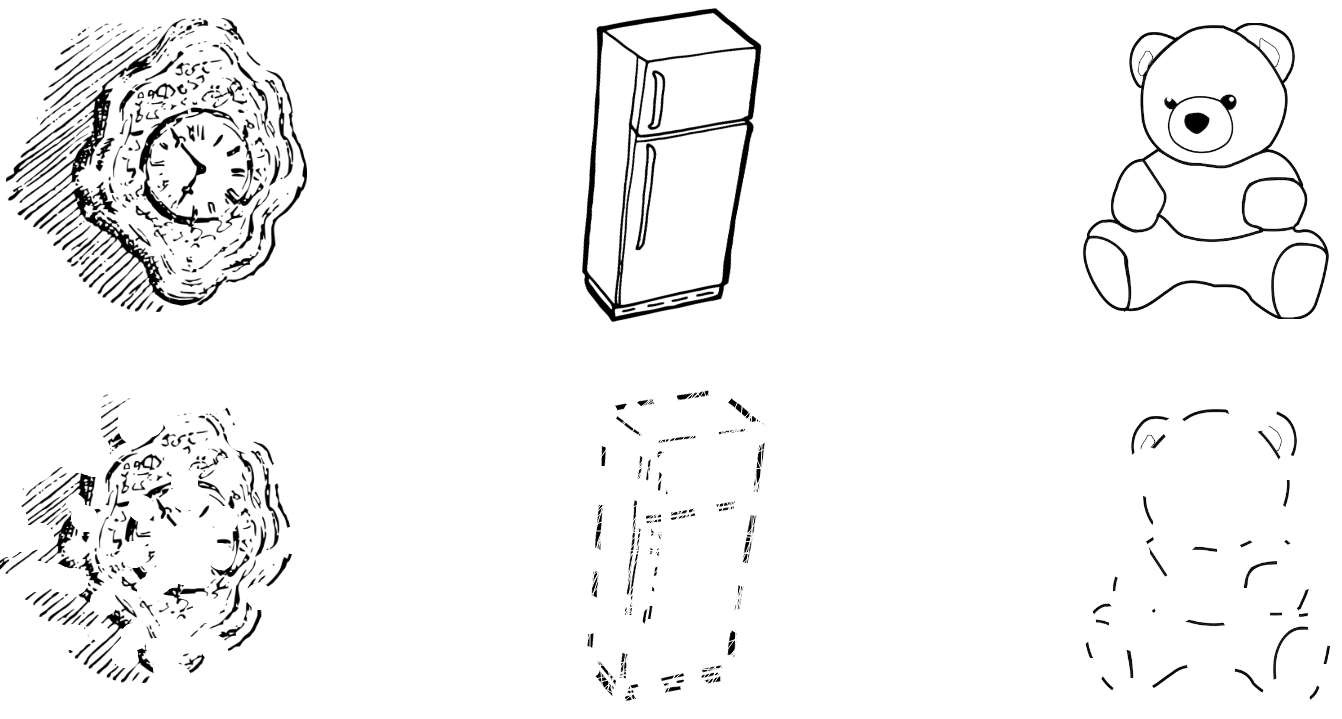}
\caption{Complete and partially complete line drawings for visually-cued subkey priming based on user subkeys from the Memory Palace.}
\label{fig:partial}
\end{figure}

We now briefly explore the act of using partial images as visual cues (figure \ref{fig:partial}) for password-subkey retrieval.  We define $p_i$ as the probability with which a random user correctly identifies a partial image such as above, when they are primed on the original completed image $i$. $n_i$ is the probability that they identify the partial image if they are not primed on the original image.

The priming effect \footnote{All images have demonstrably high priming ``strength'' \cite{schacter1993implicit} i.e. our images are already embedded in the user's mind (familiar places that they can navigate mentally)} is $\alpha$, with $\alpha > 0$ and $p_i \geq n_i + \alpha$ i.e. the probability of correctly identifying partial images with priming is greater than the probability of correct identification without priming~\cite{denning2011exploring}. Users may choose to use cues for all of their accounts, which would have required 130 cue-subkey associations for the average user last year. \cite{digital} However, this is unlikely and most users may deploy hash functions and cues for only the most sensitive data. 

What remains then, is to evaluate the success of an adversarial (without cue-subkey associations) attack.
Fortunately, this is well-established in  neuroscientific literature; we paraphrase~\cite{denning2011exploring}: \textit{Assuming an adversary knows $p_i, n_i$, and the correct label for that image, an optimal adversarial strategy is to maximize the probability of recovery of those images without knowledge of the set $U$ on which the user was primed (since this set $U$ exists uniquely in the mind of each user). The best strategy is to label each image correctly at random. However supposing an adversary is allowed to recover a user's [password] with probability at most 0.5\% (false positive rate). For valid recovery to succeed at least 97.5\% of the time (false negative rate of 2.5\%), a user would need to correctly label 135 images without prior knowledge to recover a word.} \footnote{See \cite{denning2011exploring} for a detailed proof.}

Users surveyed during a 2004 survey on Password Memorability and Security~\cite{yan2004password} were observed to use their own password generation methods, which were usually weak, yet met the security requirements demanded by websites. 
We thus propose that exploiting users' unique configurations of memory as a source of randomness enables compelling, secure password generation.

\section{Password Security Advice} \label{sec:advice}

There are three common password-hygiene errors~\cite{CommonPasswords} -- choosing simple passwords ($123456$, $iloveyou$, $qwerty$, etc.), insecure storage, and password reuse. Attacks\footnote{Cracking means an adversary with access to password hashes, has found a collision.} include guessing (common passwords), brute force, and dictionary attacks. The passwords mentioned above have 28, 40, and 32 bits of entropy respectively, which require around half a million attempts~\cite{randomness_requirement} to crack. (In reality, a hacker would guess common passwords first, and thus break these easily.) With the aid of GPU supported tools like Hashcat, Rainbow Crack etc., a 9-character password can be cracked in an alarmingly short time~\cite{Pswd_auth} -- around 18 minutes to check salted hashes for every 9-character password, assuming ideal conditions\footnote{In practice, the time taken to find a password's hash depends on the alphabet used, degree of parallelization, hardware specifications such as processor flops, etc.~\cite{hashing_time}}.

Given these issues, many websites/applications suggest strategies users should follow to create secure passwords. To better understand such password advice, we surveyed 400 highly visited platforms, compiled manually and through public lists \cite{hardwick_2020,wiki:googleplay,appleios,alexa}. Of these, 54 offered password advice; see table~\ref{table:advice} for a summary.

\begin{table}
  \centering
  \begin{tabular}{l r r}
    & \multicolumn{1}{c}{\small{\textbf{Summary of Password Advice}}} \\
    \cmidrule(r){1-2}
    {\small\textit{Parameters Suggested}}
    & {\small \textit{\% of platforms}} \\
    \midrule
    Length ($<$ 6 characters) & 20\%    \\
    Length ($>=$ 6 characters) & 20\%    \\
    Length ($>=$ 8 characters) & 41\%   \\
    Length ($>=$ 10 characters) & 19\%  \\
    Numerals  & 83\%        \\
    Uppercase  & 65\%        \\
    Special Characters  & 63\% \\
    Password Managers  & 9\%
  \end{tabular}
  \caption{Advice from 400 highly-visited websites and apps (54 provided advice).}
  \label{table:advice}
\end{table}

Websites suggest tactics such as intentionally misspelling words, replacing letters (`@' for `a', `\$' for `s', etc., so that `its raining cats and dogs'  become `1tsrAIn1NGcts\&DGS!'). However, there exist various dictionaries of special characters, common misspellings, and symbol substitutions. Hence, such tricks are ineffective against modern hackers~\cite{schneier}. An attack on with these dictionaries exposed hash collisions such as  ``Apr!l221973," and ``Qbesancon321".

What, then, is a secure password? The RSA challenge by \textit{RSA Laboratories}~\cite{reqd_security} issued random keys from 40 upto 128 bits with ciphertexts. Distributed.net has been working on the 72 bit key for over 6400 days as of July, 2020~\cite{rsa_distb}; at this pace, it takes around 200,000 days to search the entire keyspace. Currently, 72 bits of entropy  provide sufficient security; 80 bits of entropy are recommended for long-term security ~\cite{reqd_security}.

\section{Human-Computable Hashing Algorithms}\label{sec:hashing}

The functions proposed here draw upon the ideas discussed in section~\ref{sec:cogsci} to balance security and ease of use. We describe all algorithms and provide examples for cases that might otherwise be confusing. Algorithms were primarily designed to determine which approaches (subkey-generation, visualization, addition, implicit association etc.) produce the most effective and secure passwords. For this reason, they vary widely and cover a range of password generation tactics.

We perform a naive entropy calculation (assuming letter entropy values are independent) for the purposes of comparing hashing algorithms. \textit{These numbers should \textbf{not} be taken seriously as proxies for security in and of themselves, but may be useful for comparison.} Difficult-to-use schemes might push users to simply write the password down (or ignore the scheme). A ``good'' function produces high entropy passwords that are easy to compute.

Typically, hash function security is judged by pre-image resistance, collision resistance, randomness, etc.~\cite{hashFunc}. That is not easily done for our functions -- we cannot generate billions (or even millions) of hashes, as the process of generation relies on individuals' unique memory representations and sources of randomness  (discussed below and in appendix \ref{sec:appendix}). We discuss some metrics we can use in section~\ref{sec:hash-analysis} and cryptographic details in appendix~\ref{sec:appendix}.

\subsection{Description of the Schemes} \label{sec:function-descriptions}

We describe the following human-computable hash functions: Memory Palace, Scrambled Box, Song Password, Internal Sentence. $w$ is the website name, $s$ is the single secret user key, and $h$ is the candidate for $F$. $F$ and $h$ are functions of $R$, the unique configuration of each user's memory. Each source of randomness is indicated by $_R$ and specified at the end of each algorithm. Sources are elaborated on in Appendix \ref{sec:appendix}. Common sources of randomness across all algorithms: unique memory associations; choosing between symbols, numerals or letters on the same key.

\medskip

\noindent
\textbf{Memory Palace.} 
$s$: A location$_R$ very familiar to the user. $h_R(s,w)$:

\begin{itemize}
    \item[Step 1] \textit{subkey generation} Mentally navigate the location using each letter in $w$. For vowels turn left and walk straight$_R$, else turn right and walk straight. After reaching the end of the website name, think of a word (or words) that describe what the user faces.
    (If $w = gmail$, visualizing a familiar location, mentally move right and straight twice then left and straight twice, then right and straight twice. $s$ = a description of what you face.)

    \item[Step 2] \textit{group sum} Divide the word(s) into groups of 2 letters (pairs). Sum each group using letter values to create a new letter. (Letters map to $\{a=1, \dots , z=26\}$, if sum overflows, subtract 26 from the sum.) If $s$ can't be evenly split add a favorite letter$_R$ to the end.
    (If $s$ = white birds, split into wh, it, eb, ir, ds. Sum into $w+h = e$, $i+t = c$,$e+b=g$,$i+r=a$, $d+s=w$)

    \item[Step 3] \textit{group character} If the first letter of a pair is a vowel, write the symbol/letter above and to its immediate diagonal$_R$ left on the keyboard after the letter from the group sum. Else, the symbol/letter above to its immediate diagonal right on the keyboard. (Described and illustrated visually during the survey.)

    $password$: Alternate group sum and group character.
    (Alternating group sum letters with corresponding diagonal symbols, $password$ = e3cfgya1w3.)
\end{itemize}
Randomness: Spatial characteristics of direction, number of steps to take when walking.  Letter preference when appending letters to make the length of $s$ even. Interpretation of diagonal angle,  choosing the $i^{th}$ symbol along the diagonal. 

\medskip

\noindent
\textbf{Scrambled Box.} 
$global$: A 10x10 table of symbols, numbers and letters (repetitions allowed). Movements associated with each story element (can be changed): Sad = up; Memorable characters (Animals, Villains etc) = diagonal to the right and down; Events that move the story forward = horizontal to the right; Happy = move to the opposite corner of the table.

\begin{figure}[ht]
\includegraphics[width=\textwidth]{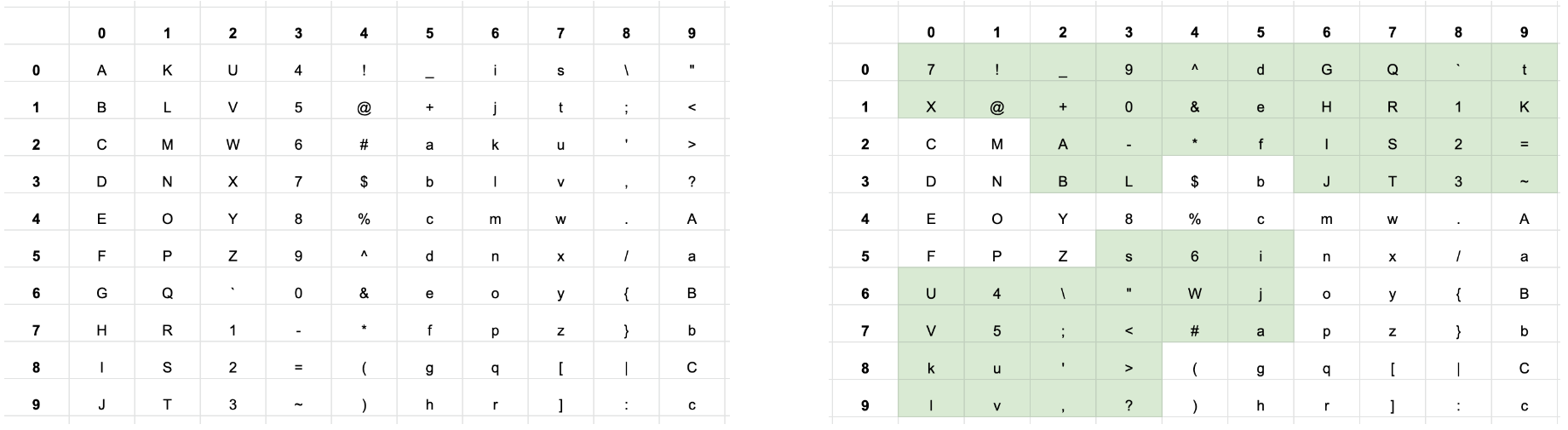}
\caption{Example 10x10 box and S-box, with scrambling highlighted}
\label{fig:merged-boxes}
\end{figure}

$s$: A well-known easily remembered story$_R$ name. $h_R(s,w)$:
\begin{itemize}
    \item[Step 1] \textit{S-box generation} Find 4 elements (e.g., emotions, events, memorable characters) in the story's plot and write them down in order. For the $x^{th}$ element of the story, choose a $x \times x$ square and move it by $x$ squares, using the associated direction. Swap it with the square it replaces.

    \item[Step 2] \textit{S-box-website mapping} Connect the story to the website to come up with a word/words$_R$. Convert letter values (mapping a=0, z=25) in the word(s) to integers, add a 0 to the number if it is a single-digit integer. Treat integers as (x,y) coordinates and find the corresponding characters in the table. Save this sequence of characters as the $password$.
    (For example: Connecting Tarzan to Amazon may result in the word ``shirt'' which maps to letter values ``19 8 9 18 20''. Adding 0s to single digits, ``19 80 90 18 20'', and mapping to the S-box results in coordinates (1,9), (8,0) etc. The password: v'tu\_)
\end{itemize}

\noindent
\textbf{Song Password.} 
This method relies on two sources of randomness -- songs and a 4 digit key. 
$s$: A 4-digit pin. 
$h_R(s,w)$:

\begin{itemize}
    \item [Step 1] Reduce $w$ to a 4 letter mnemonic.
    (\textit{Flipkart} becomes \textit{f p k t})
    
    \item [Step 2] Choose a 4 digit key$_R$. 
    (3 8 1 9) 
    
    \item [Step 3] Choose 4 songs$_R$ starting from each letter of the mnemonic. These should be songs (not necessarily in English!) that have significance or are easy to remember. 
    (\textit{Fade}, \textit{Panama}, \textit{King of Mars} and \textit{Teddy Boy}.)
    
    \item [Step 4] Choose words$_R$ from each song, corresponding to each digit of the key, and concatenate to form a \textit{Song String}, $S_x$. 
    ($3^{rd}$ word from \textit{Fade}, $8^{th}$ word from \textit{Panama}, $1^{st}$ word from \textit{King of Mars} and $9^{th}$ word from \textit{Teddy Boy}.)
    
    \item[Step 5] After every vowel in $S_x$, insert a special character closest$_R$ to the vowel on the keyboard. If there is more than 1 special character equidistant from the vowel, choose$_R$ one and remember it. (For $o$, `(' or `)', for $e$ `\$' or `\#'.)
    
    \item[Step 6] Choose three characters$_R$ (letters or symbols) and move them to the end of the password. Repeat with another group of three. Then remove every alternate character (starting with the first). $password$: resultant string. 
\end{itemize}
Sources of randomness: Interpretations of linguistic fillers as words, choice of special character and characters to move. 

\medskip

\noindent
\textbf{Internal Sentence.} 
$s$: A rarely used word$_R$ from any language. $h_R(s,w)$: Create a sentence connecting the website to the word. $password$: Sentence created.

\section{Analysis of Hash Functions}\label{sec:hash-analysis}

This section analyzes the security and real-world effectiveness of our hash functions via several metrics, including a user study: 134 individuals aged 18-25 were surveyed, with each user generating passwords using 2 different randomly-assigned algorithms. Each algorithm had an average of 56 responses. We also include Cue-Pin-Select~\cite{CuePinSelect} in our survey.

\subsection{Generation and Retention} \label{sec:generation_retention}

Previous attempts have suggested ``intolerably slow'' methods~\cite{denning2011exploring}. Our protocols can be executed by the average user within 5 minutes for generation, and recollection time decreases significantly with repetition. The key human-computability properties of $F_R$ are: (1) Reliance on cognitive and visual cues for stable, rapid recall\footnote{Some of which are proven to last in memory 17 years without repeated rehearsal~\cite{denning2011exploring}} (2) Minimal effort, and limited access to education or writing resources.

Some of our methods retain significant security without access to any external materials for generation. The Memory Palace and Internal protocols need only a keyboard (or pictures of standard keyboards; no writing materials or internet, though access to these would decrease cognitive load).

\begin{figure}[ht]
\centering
\includegraphics[width=0.7\textwidth]{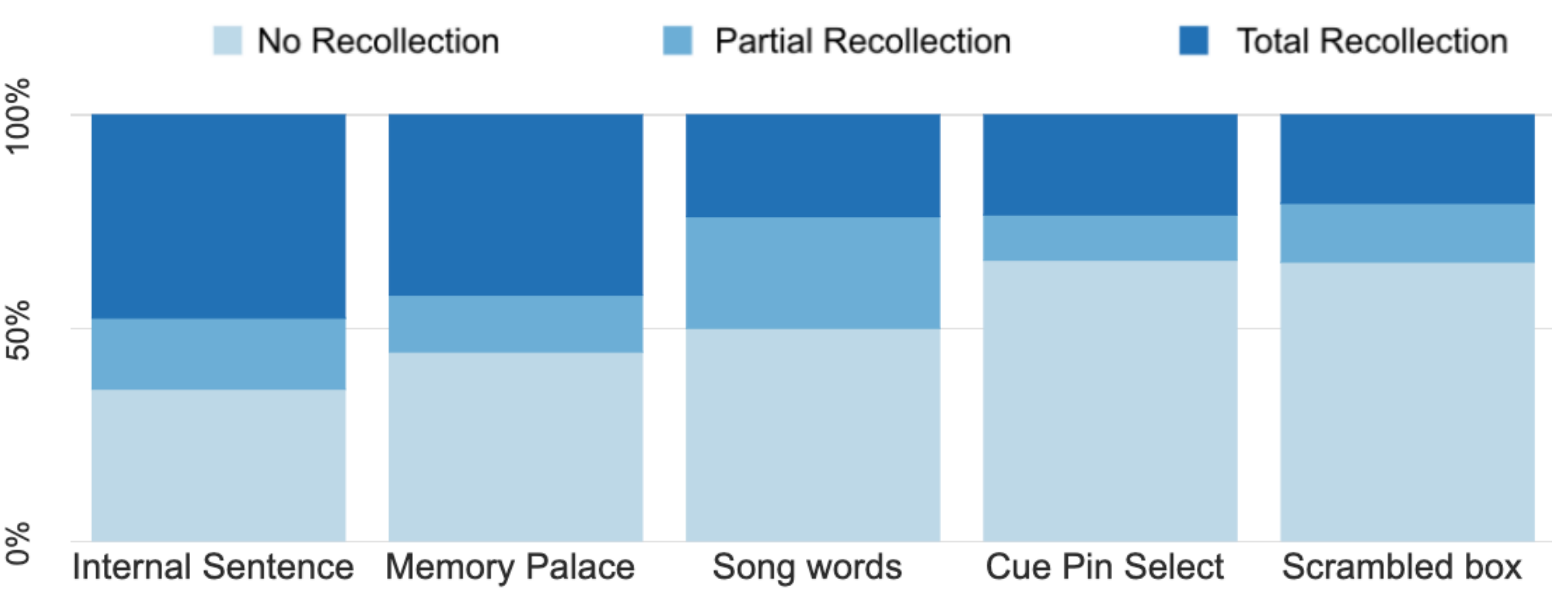}
\caption{Password recollection visualized (based on table~\ref{tab:table1})}\label{fig:recall}
\end{figure}

The ability to recall or regenerate a password is essential to its effective security; lower memorability leads to frequent passwords resets and frustration that may lead to users abandoning the algorithm. Users were surveyed over a week to test password retention. See fig.~\ref{fig:recall} and table~\ref{tab:table1}. Methods with less successful recall (Cue-Pin-Select, Song password and Scrambled box) seem to require more explicit memorization. Associative techniques can exponentially increase ease of password recollection (Memory Palace, Internal Sentence), and provably improve system security~\cite{chakravarthy2011novel}. Therefore we recommend the use of partial visual cues for subkey association whenever possible.

\begin{figure}[ht]
\centering
\includegraphics[width=0.9\textwidth]{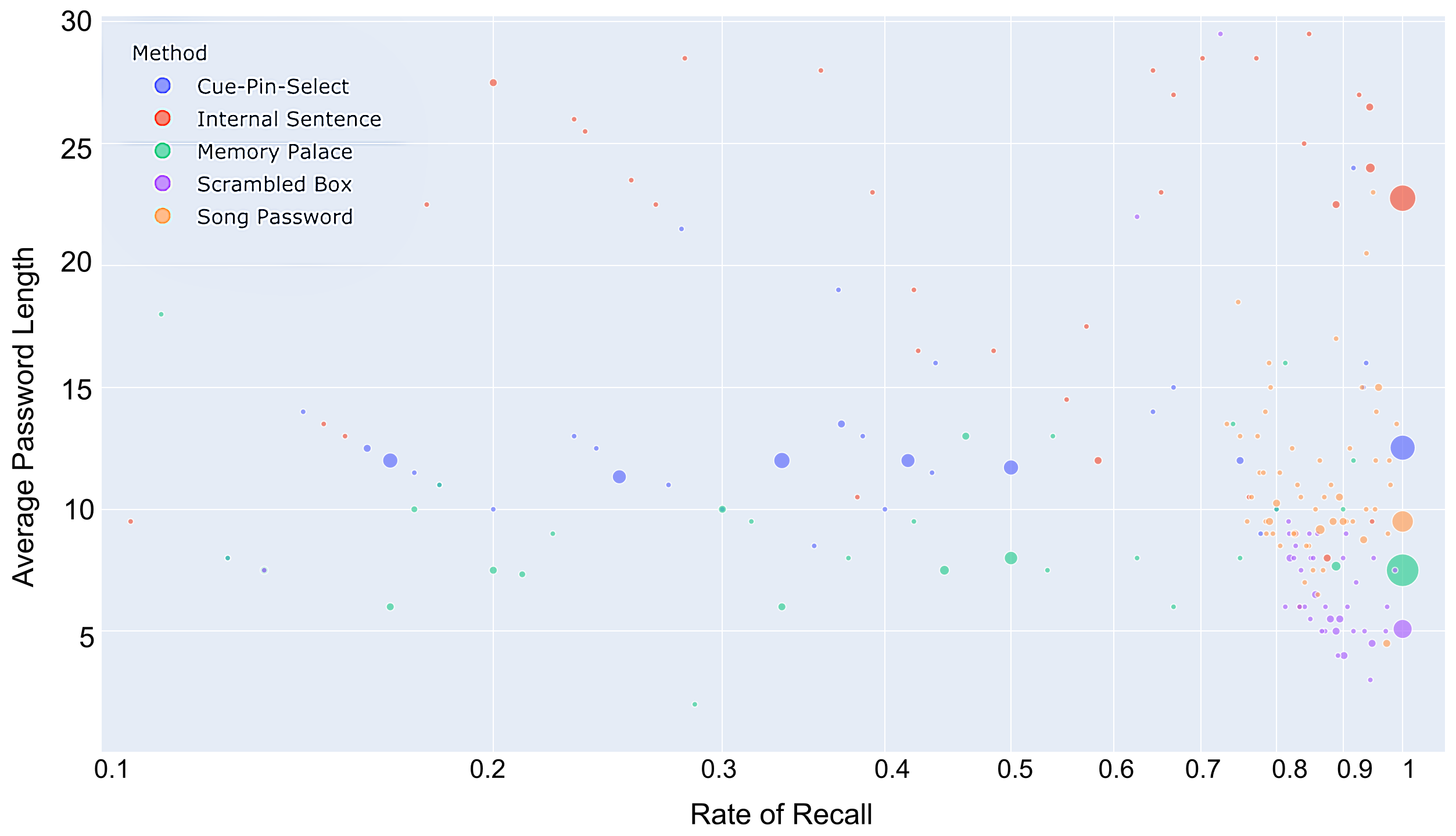}
\caption{Bubble chart of the rate of password recollection for hash functions.\newline Each function is represented by a color; the frequency of each rate of recall (recall measured by: $\mathbb{S}(p_i,p_r)$ where $\mathbb{S}$ corresponds to the Gestalt Pattern Matching (Ratcliff/Obershelp string similarity algorithm \cite{paul2004black,python}) corresponds to the size of each bubble, $p_i$ is the initial password and $p_r$ is the remembered password; the axes measure password length and the frequency of each length.}
\label{fig:bubble}
\end{figure}

The rightmost area of figure~\ref{fig:bubble} indicates perfectly recalled passwords, with larger bubbles indicating a more significant percentage of users with perfect recollection. Ideal functions are large bubbles at the rightmost end of the graph with an average password length above 10 characters (see section~\ref{sec:advice}).

\begin{table}
  \centering
  \begin{tabular}{l r r r}
    & & \multicolumn{2}{c}{\small{\textbf{Password Memorability}}} \\
    \cmidrule(r){3-4}
    {\small\textit{Hashing algorithm}}
    & {\small \textit{Attempts}}
    & {\small \textit{Complete R}}
    & {\small \textit{Partial R}} \\
    \midrule
    Internal Sentence & 42 & 21 (50\%) & 7 (17\%) \\
    Memory Palace & 45 & 19 (43\%) & 6 (14\%) \\
    Song words & 42 & 10 (24\%) & 11 (27\%) \\
    Cue Pin Select & 47 & 11 (24\%) & 5 (11\%) \\
    Scrambled box & 29 & 6 (21\%) & 4 (14\%) \\
  \end{tabular}
  \caption{R: recall/regeneration of passwords. Attempts: Number of people who attempted R. Total R: exact recall/regeneration of 1 or more passwords created. }~\label{tab:table1}
\end{table}

Each time a password is recalled using a key, a user-familiar memory (object, space, color etc) is associated with the key. This key-memory association is repeated until thinking of one automatically brings the other to mind~\cite{loterre}. We emphasize that, as in all reasonable systems, the generation method is public, and the only secret that needs to be remembered is this key.

The advantage of involving the methods proposed in this paper (such as visual, associative, implicit memory) is that they can be adapted to existing password generation methods. \textit{E.g.}, Cue-Pin-Select can be modified to choose random words with visual or associative cues drawing on implicit memory.

\subsection{Effective Security}\label{sec:analyse_effective}

\begin{figure}[ht]
\centering
  \includegraphics[width=0.8\textwidth]{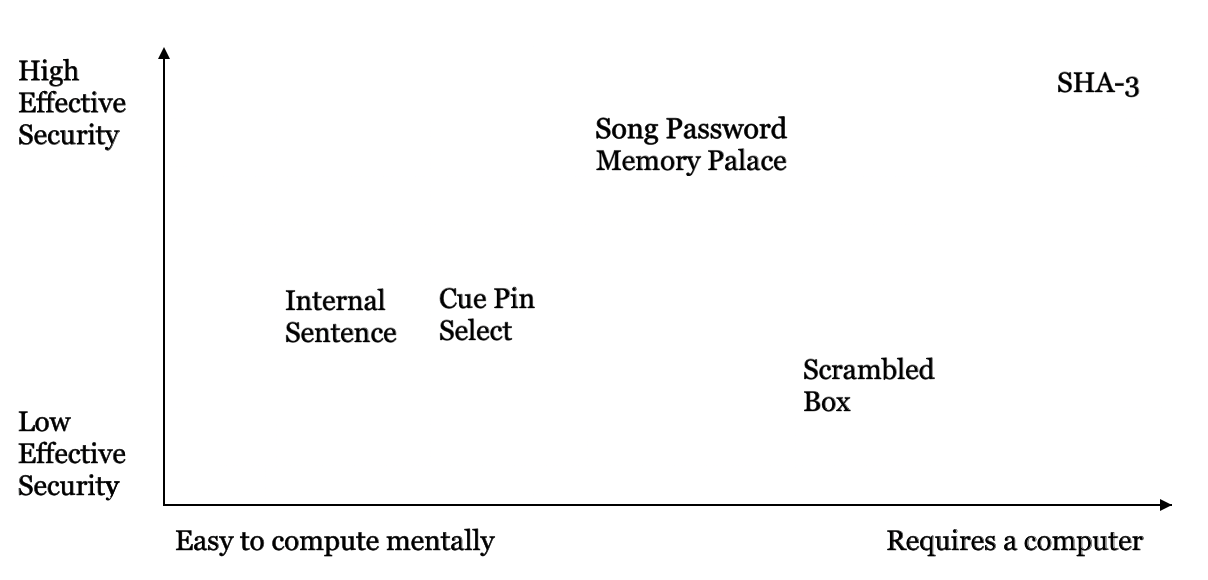}
  \caption{Mapping effective security (password security and user comfort with algorithms) and ease of use (user perception on a scale of requiring no resources, to requiring computers). \textit{Axes are exaggerated subjectively for illustration.}}~\label{fig:es}
\end{figure}

We propose the concept of \textit{effective security}. A password generation scheme may be incredibly secure, but is useless\footnote{Assuming an appropriate threat actor -- imagining an adversarial `evil' sibling with occasional read-only access to your living space is a useful rule of thumb.} if it is so hard that most users just write down their passwords. (See fig.~\ref{fig:es}.)
The effective security of a function $F_R$ is the actual difficulty of breaking one of its assumptions in real-world use by laypersons. 
The ideal human-computable hash function is easy enough (and grows easier through repeated use) to encourage humans to use it, while retaining the necessary entropy to ensure security by resisting attacks.

\begin{figure}[ht]
\centering
\includegraphics[width=\textwidth]{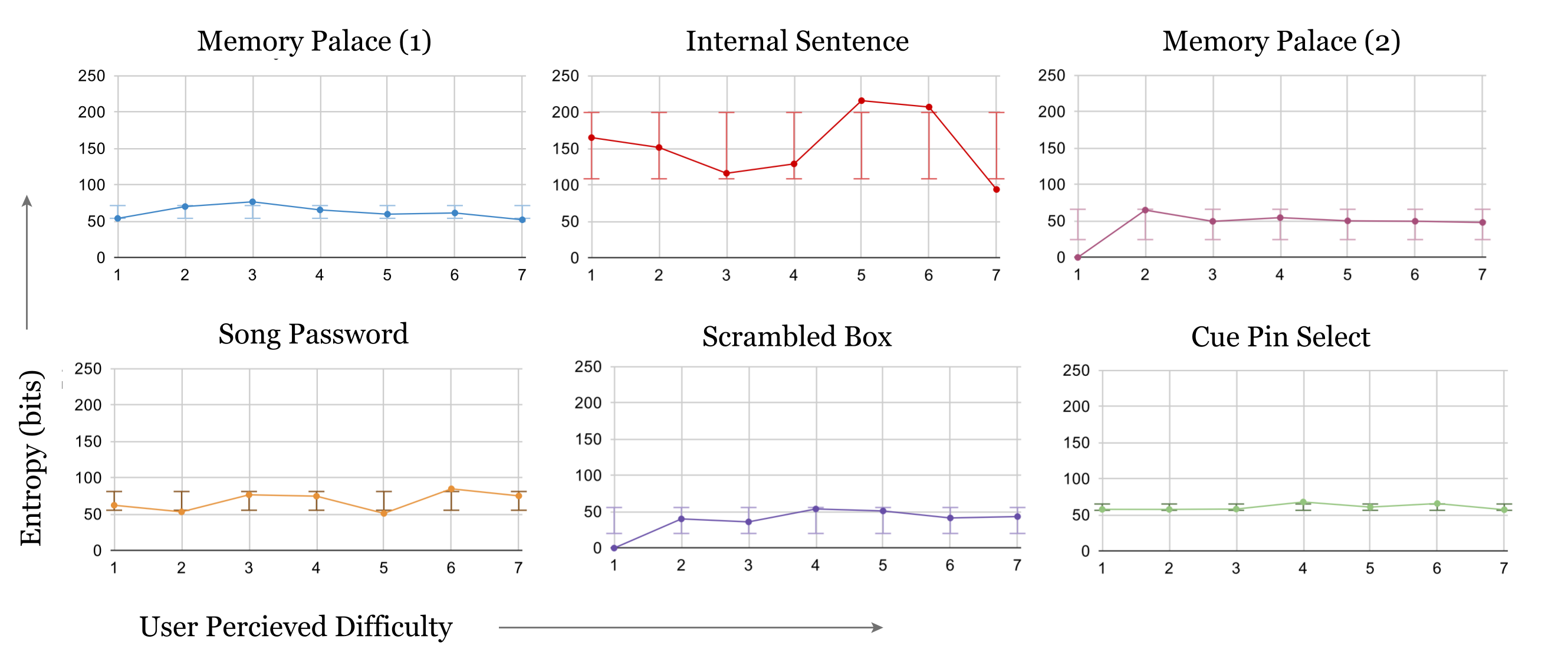}
\caption{Each point represents the mean entropy of passwords with some user perceived difficulty (std. dev. error bars). Memory Palace Step 1 was presented as ``method 1'' to users; 2 included all steps described in section~\ref{sec:function-descriptions}. X-axis: User Perceived Difficulty; Y-axis: Password Entropy.}
\label{fig:diffvsent1}
\end{figure}

Traditional cryptographic evaluations are built to evaluate functions designed for computers. We present a range of strategies for security evaluation in appendix~\ref{sec:appendix}. These strategies are \textbf{not} indicative of security by themselves, but taken in combination provide a good measure of the relative security of each function; further work is required to understand the security of such methods.

\subsection{User Study and Improvements}\label{sec:subuserstudy}

\begin{figure}[htbp]
    \centering
    \includegraphics[width=0.8\textwidth]{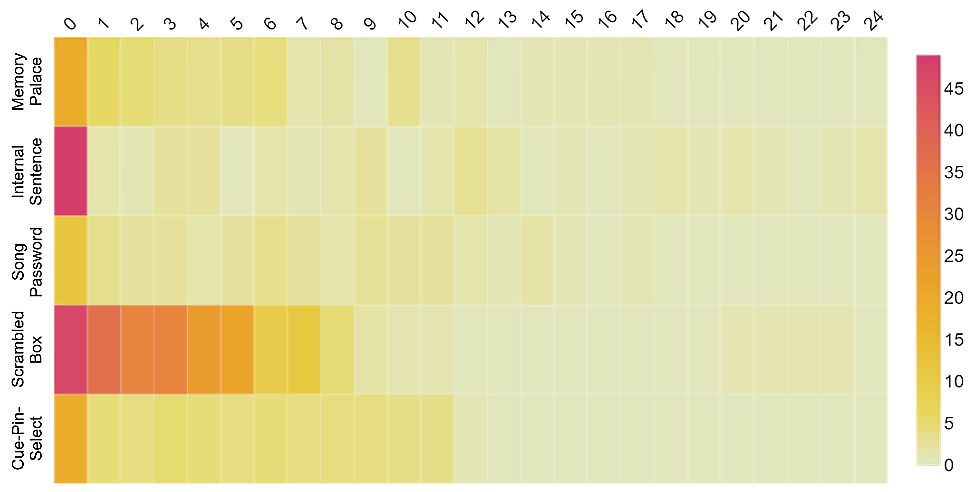}
    \caption{Heatmap of the incidence of capital letters at different indexes. Passwords $>25$ characters are omitted}
    \label{fig:capitalletters2}
\end{figure}

We perform a survey comprising n=134 individuals, with an average of 56 users suggesting improvements for each algorithm. We present baseline entropy evaluations for each function\footnote{Assuming character entropies are independent. We do not consider dictionary attacks, character frequencies etc as these would require a large number of passwords to be statistically valid, and due to unique user memory configurations $R$ we cannot computationally generate large numbers of passwords.}, measure passwords from each function against current security standards and suggest improvements based on user feedback.

Our human computable hash functions average a password entropy of 78.07 bits, significantly higher than the average entropy of 40.54 bits per password as estimated by Microsoft~\cite{florencio2007large}. These functions also encourage higher entropy by increasing use and distribution of symbols and capitalization. Memory Palace, Song Password, and Scrambled Box increase the number of symbols per average password to 3.188 symbols, compared to a baseline of 0.2 symbols. Capital letters decrease to 0.412, lower than 1.1 without hash functions. However, as evident in figure~\ref{fig:capitalletters2}, capitalization is more distributed across location, rather than concentrated towards the first character of the password~\cite{jonathan_2019}.


Our results are reasonably representative of the general population of password users~\cite{mazurek2013measuring}. Our choice of sample size is based on \cite{komanduri2011passwords} and \cite{mazurek2013measuring}. Our sample is drawn from students in a medium-sized university in India and may be applicable to similar demographic profiles. In addition, the sample represents a range of language, educational and income backgrounds. However, the proportions of these demographics are not the same as the general population. Beyond the obvious age bias (college-aged individuals), the sample is biased towards individuals willing to participate in the survey in exchange for food and money (both standardized), and all data is self-reported. In addition, Cochran's formula \cite{kotrlik2001organizational} recommends a sample size of 100 individuals based on the proportion of internet users in the world (53.6\% of the global population in 2019 ~\cite{measuring}), a 95\% confidence interval and a 10\% error margin. Compared to previous human computable password research \cite{CuePinSelect}, we use a significantly larger sample size with a more representative demographic of password users. Thus, our results, extrapolated prudently, can apply to the broader population.

Scrambled Box and Song require writing (the latter requires access to a music repository) and are harder than the first two methods for users. Song and Cue-Pin-Select also require greater intermediate key generation -- choosing and explicitly recalling random unique words/songs and a pin/word. Comparatively, Internal Sentence and Memory Palace use associations already familiar to users.

\begin{table}
  \centering
  \begin{tabular}{l r r r}
    & & \multicolumn{2}{c}{\small{\textbf{Graceful degradation and Entropy}}} \\
    \cmidrule(r){3-4}
    {\small\textit{Function}}
    & {\small \textit{Mean entropy (bits)}}
      & {\small \textit{Standard Deviation}}
    & {\small \textit{Graceful Degradation}} \\
    \midrule
    Internal Method & 153.95 & 97.14 & 0.66   \\
    Memory Palace & 51.08 & 25.84 & 0.38  \\
    Song Words & 74.57 & 44.12 & 0.49 \\
    Cue Pin Select & 61.96 & 17.62 & 1.06 \\
    Scrambled Box & 45.15 & 33.15 & 0.84 \\
  \end{tabular}
  \caption{Graceful degradation, mean entropies, and their standard deviations.}~\label{tab:gracefuldeg}
\end{table}

Graceful degradation in table \ref{tab:gracefuldeg} measures increase in difficulty with decrease in education levels. Larger graceful degradation corresponds to functions that require higher education levels. 

\textbf{Memory Palace.}
With the aid of partial visual cues, memorizing hundreds of cues for subkey generation (objects, areas, memories etc) \cite{digital} is unnecessary and the user can focus on subkey-cue associations. Most keys were in 4\% of the 100 most common words in English, including references to common household objects and local languages. After hashing subkeys with each website, no English words were identifiable (excluding users who misinterpreted instructions). 

Users were satisfied with the security but suggested clearer navigational guidance. A common struggle was navigating dead-ends with visually unremarkable cues. A significant proportion of users struggled with Step 2 and favored Step 1 and 3. 
Some users stated they would adapt Step 1 for future password generation.

\noindent
\textbf{Scrambled Box.}
The key is the `box' of pseudo-randomly scrambled symbols. This can be written down or shared, but must be unique to each user, who only needs to remember website-subkey associations. Users found rearranging symbols hard and preferred fewer instructions, but liked the lack of memorization.

\begin{table}
  \centering
  \begin{tabular}{l r r r}
    & & \multicolumn{2}{c}{\small{\textbf{Survey Results}}} \\
    \cmidrule(r){3-4}
    {\small\textit{}}
    & {\small \textit{password length}}
      & {\small \textit{security}}
    & {\small \textit{difficulty (1-7)}} \\
    \midrule
    Internal Sentence & 25.91 & 9.90 & 2.52  \\
    Memory Palace & 8.42 & 86.06 & 5.38 \\
    Song Words & 11.50 & 92.16 & 5.41 \\
    Cue Pin Select & 12.29 & 3.21 & 4.44 \\
    Scrambled Box & 6.71 & 94.11 & 5.68 \\
  \end{tabular}
  \caption{Security refers to the \%age of passwords with $\geq$1 Number or Symbol. Length and difficulty are averages. Difficulty was assessed by users.}~\label{tab:table2}
\end{table}

\noindent
\textbf{Song Password} 
This scheme amplifies randomness in the input. For example, using songs: \textit{Fade}, \textit{Panama}, \textit{King of Mars}, \textit{Teddy Boy} with user one’s PIN$_1$ = 3819 and user 2’s PIN$_2$ = 7144, passwords generated are \texttt{mse\$i(o)*} and \texttt{tsto)mhS} (a similarity of 0.33\% \cite{paul2004black,python}).

Users struggled with pins and associating different songs. 
Some users preferred not to remove or shift alternate characters, while others remarked they would adapt this method for future password generation.

\noindent
\textbf{Internal Sentence.}
Users preferred this method for ease of use but struggled with remembering word order, verb and adjective choice, etc. or found passwords generated too long to recall. Users felt this method was insecure as it did not generate special characters or capitalization. The entropy for this method is misleading, as passwords often contain words susceptible to dictionary attacks.

\noindent
\textbf{Cue-Pin-Select}
Word and pin recollection were challenging, users preferred associating words with cues over random cues, and suggested reducing the number of random words from 6 to 4. In general users requested stronger associative and implicit memory modifications to the method. Across all passwords and algorithms mentioned in \ref{sec:function-descriptions}, average password entropy is 78.07 bits and average password length is 11.83 characters (i.e. numbers, symbols and letters).

\begin{figure}[htbp]
    \centering
    \includegraphics[width=\textwidth]{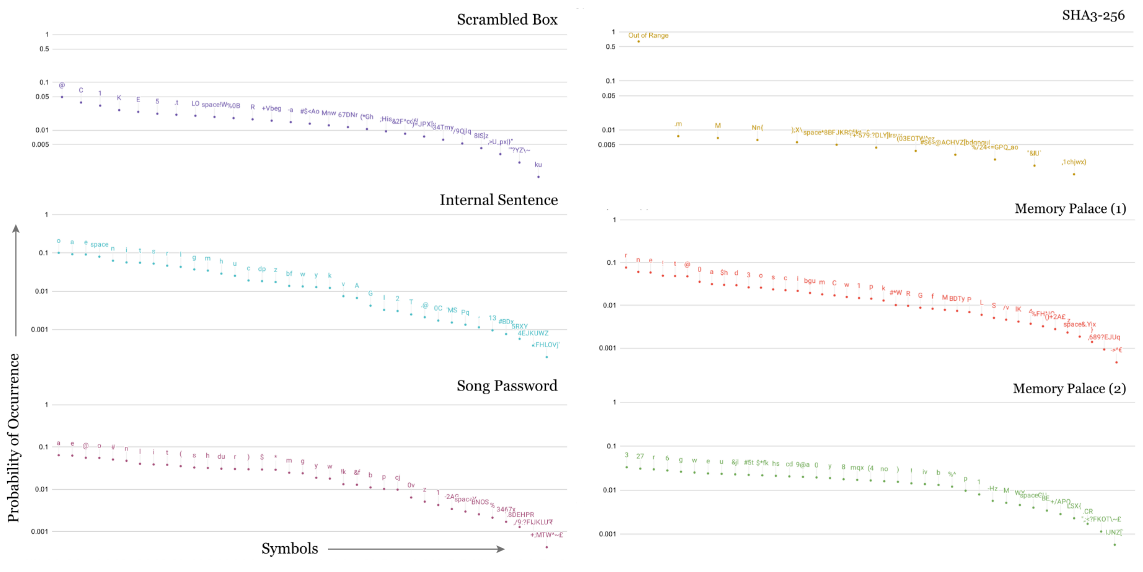}
    \caption{Symbol occurrence by method. Y-axis: log scale. X-axis based on symbol rank (most to least probable). SHA3-256 hashes were converted to latin-1 encoding to get typable character frequencies~\cite{githubcode}. Memory Palace as in figure~\ref{fig:diffvsent1}.}
    \label{fig:symbolsheatmap}
\end{figure}

\begin{table}
  \centering
  \begin{tabular}{l r r r}
    & & \multicolumn{1}{c}{\small{\textbf{Testing Accuracy (in \%)}}} \\
    \cmidrule(r){2-3}
    {\small\textit{Scheme}}
    & {\small \textit{100 epochs}}
    & {\small \textit{200 epochs}} \\
    \midrule
    Internal Method & 53.13 & 58.41   \\
    Memory Palace & 18.42 & 19.91   \\
    Song Words & 21.71 & 23.37 \\
    Cue Pin Select & 47.56 & 46.76  \\
    Scrambled Box & 29.31 & 28.44 \\
  \end{tabular}
  \caption{We used a Long Short Term Memory Network~\cite{lstm_adv} to learn dependencies. The 50-cell LSTM was tested with two trials of 100 and 200 epochs.}~\label{table:ml}
\end{table}

\subsection{Machine-Learning Based Analysis using LSTMs}\label{sec:ml_analysis}

A simple machine learning system was used to predict the $k^{th}$ character given the previous $k-1$ characters of the password to further evaluate randomness. This is based on a long line of research starting from Shannon's entropy experiment~\cite{shannon}. 
We used all characters except the last for training (see  Table~\ref{table:ml}).


\section{Real-World Password Generation Methods}\label{sec:user_methods}

How do people currently generate (and remember) passwords? Our survey suggests that people use a combination of words, followed by digits and symbols (in that order), indicating construction in order of ease of recollection.
Common associations: names of relatives, fictional characters, nicknames, etc.; digits or symbols — birth dates, reversed phone numbers, even credit card numbers!

Some used inventive techniques to balance security with memorability: account expiry dates, rhymes, snacks and manufacturing dates, and slang words. Several users reused passwords with the awareness of compromised security, citing a lack of convenient options. A small population added random words from different languages. (Full database of results omitted for brevity.)

We observed that users designed passwords with human adversaries in mind and thus mistakenly believed that using animals or objects they disliked, using common character substitutions for letters (``leetspeak''), or misspelling words created a secure password. Based on previous work~\cite{6805770} and our survey, we recommend all platforms with password requirements brief users on current strategies used by computationally-equipped adversaries, such as dictionary attacks, frequency analysis etc. to reduce the usage of insecure passwords.

\section{Conclusion}

We propose a range of human-computable hashing algorithms with string and non-string inputs, designed for password generation and management. We exploit users' unique memory configurations to drive our design, drawing upon existing neuroscientific research. We also collate current password advice across hundreds of popular websites and applications, and survey users on their current password generation methods, highlighting major issues and discussing mitigation.

Our functions are validated and tested using a survey ($n=134$) to understand real-world usability. We note that larger surveys across a range of age groups are required to better classify the security and usability implications. Further work also needs to be done to explore the kinds of atomic human-computed operations that produce stable output useful for cryptography.



%
%
%
\bibliographystyle{splncs04}
\bibliography{references}

\begin{thebibliography}{10}
\providecommand{\url}[1]{\texttt{#1}}
\providecommand{\urlprefix}{URL }
\providecommand{\doi}[1]{https://doi.org/#1}

\bibitem{alexa}
Alexa: The top 500 sites on the web, \url{https://www.alexa.com/topsites}

\bibitem{baddeley1997human}
Baddeley, A.D.: Human memory: Theory and practice. psychology press (1997)

\bibitem{bestreviews_2018}
BestReviews: Which password managers have been hacked? - best reviews (Jul 2018), \url{https://password-managers.bestreviews.net/faq/which-password-managers-have-been-hacked/}

\bibitem{CuePinSelect}
Blanchard, N., Gabasova, L., Selker, T., Sennesh., E.: {Cue-Pin-Select, a Secure and Usable Offline Password Scheme.} ffhal-01781231  (2018)

\bibitem{blocki2013naturally}
Blocki, J., Blum, M., Datta, A.: Naturally rehearsing passwords. In: International Conference on the Theory and Application of Cryptology and Information Security. pp. 361--380. Springer (2013)

\bibitem{blocki2014towards}
Blocki, J., Blum, M., Datta, A., Vempala, S.: Towards human computable passwords. arXiv preprint arXiv:1404.0024  (2014)

\bibitem{measuring}
Bogdan-Martin, D.:  (2019), \url{https://www.itu.int/en/ITU-D/Statistics/Documents/facts/FactsFigures2019.pdf}

\bibitem{hashing_time}
Buys, B.: Estimating password crack times, \url{https://www.betterbuys.com/estimating-password-cracking-times/}

\bibitem{chakravarthy2011novel}
Chakravarthy, A., Raja, P.V.K., Avadhani, P., et~al.: A novel approach for password authentication using bidirectional associative memory. arXiv preprint arXiv:1112.2265  (2011)

\bibitem{300billionpasswords}
Cybersecurity~Ventures, C.M.: New report finds 300 billion passwords will be at risk by 2020. Available at \url{https://cybersecurityventures.com/300-billion-passwords/} (2017)

\bibitem{denning2011exploring}
Denning, T., Bowers, K., Van~Dijk, M., Juels, A.: Exploring implicit memory for painless password recovery. In: Proceedings of the SIGCHI Conference on Human Factors in Computing Systems. pp. 2615--2618 (2011)

\bibitem{randomness_requirement}
Eastlake, Schiller, C.: Randomness requirements for security (Jun 2005), \url{https://tools.ietf.org/pdf/rfc4086.pdf}

\bibitem{florencio2007large}
Florencio, D., Herley, C.: A large-scale study of web password habits. In: Proceedings of the 16th international conference on World Wide Web. pp. 657--666 (2007)

\bibitem{hashFunc}
Fung, E.: Hash functions, \url{https://www.cs.usfca.edu/~ejung/courses/686/lectures/05hash.pdf}

\bibitem{gedeon_2020}
Gedeon, K.: Popular password managers can get hacked: Should you keep using them? (Mar 2020), \url{https://www.laptopmag.com/news/popular-password-managers-can-get-hacked-should-you-keep-using-them}

\bibitem{harrispoll}
Google, H.P.s.: Online security survey google / harris poll. Available at \url{http://services.google.com/fh/files/blogs/google_security_infographic.pdf} (Feb 2019)

\bibitem{digital}
Guardian, D.: Uncovering password habits: Are users' password security habits improving? (infographic) (Dec 2018), \url{https://digitalguardian.com/blog/uncovering-password-habits-are-users-password-security-habits-improving-infographic}

\bibitem{shannon}
Hamid~Moradi, J.W.G.B.: Entropy of english text (1998), \url{http://citeseerx.ist.psu.edu/viewdoc/download?doi=10.1.1.92.5610&rep=rep1&type=pdf}

\bibitem{hardwick_2020}
Hardwick, J.: Top 100 most visited websites by search traffic (as of 2020) (May 2020), \url{https://ahrefs.com/blog/most-visited-websites/}

\bibitem{jonathan_2019}
Jonathan: Beyond password length and complexity (May 2019), \url{https://resources.infosecinstitute.com/beyond-password-length-complexity/#:~:text=Password Length,numbers and 0.2 special characters.}

\bibitem{komanduri2011passwords}
Komanduri, S., Shay, R., Kelley, P.G., Mazurek, M.L., Bauer, L., Christin, N., Cranor, L.F., Egelman, S.: Of passwords and people: measuring the effect of password-composition policies. In: Proceedings of the sigchi conference on human factors in computing systems. pp. 2595--2604 (2011)

\bibitem{kotrlik2001organizational}
Kotrlik, J., Higgins, C.: Organizational research: Determining appropriate sample size in survey research appropriate sample size in survey research. Information technology, learning, and performance journal  \textbf{19}(1), ~43 (2001)

\bibitem{loterre}
Loterre: \url{https://www.loterre.fr/skosmos/P66/en/page/-SQ2MHWHN-Q}

\bibitem{mazurek2013measuring}
Mazurek, M.L., Komanduri, S., Vidas, T., Bauer, L., Christin, N., Cranor, L.F., Kelley, P.G., Shay, R., Ur, B.: Measuring password guessability for an entire university. In: Proceedings of the 2013 ACM SIGSAC conference on Computer \& communications security. pp. 173--186 (2013)

\bibitem{o'flaherty_2019}
O'Flaherty, K.: Password managers have a security flaw -- here's how to avoid it (Feb 2019), \url{https://www.forbes.com/sites/kateoflahertyuk/2019/02/20/password-managers-have-a-security-flaw-heres-how-to-avoid-it/}

\bibitem{paul2004black}
Paul, E.: Black. 2004. Ratcliff/obershelp pattern recognition. Dictionary of Algorithms and Data Structures  \textbf{17} (2004)

\bibitem{press}
Press, O.U.: The oxford 3000, \url{https://www.oxfordlearnersdictionaries.com/about/oxford3000}

\bibitem{python}
Python Software~Foundation, P.S.F.: 7.4. difflib - helpers for computing deltas (2020), \url{https://docs.python.org/2/library/difflib.html}

\bibitem{githubcode}
Ruthu, R.: Github, code reference (Jul 2020), \url{https://github.com/debayanLab/trenchcoat}

\bibitem{SplashData}
Safe, S.: \url{https://splashdata.com/press/releases.htm}

\bibitem{schacter1993implicit}
Schacter, D.L., Chiu, C.Y.P., Ochsner, K.N.: Implicit memory: A selective review. Annual review of neuroscience  \textbf{16}(1),  159--182 (1993)

\bibitem{schneier}
Schneier, B.:  (2014), \url{https://www.schneier.com/blog/archives/2014/03/choosing_secure_1.html}

\bibitem{lstm_adv}
{Shi}, Z., {Shi}, M., {Li}, C.: The prediction of character based on recurrent neural network language model. In: 2017 IEEE/ACIS 16th International Conference on Computer and Information Science (ICIS). pp. 613--616 (2017)

\bibitem{smith_2020}
Smith, A.: Americans, password management and mobile security (Aug 2020), \url{https://www.pewresearch.org/internet/2017/01/26/2-password-management-and-mobile-security/}

\bibitem{rsa_distb}
Stats, D.: Rsa challenge (Jun 2020), \url{http://stats.distributed.net/projects.php?project\_id=8}

\bibitem{appleios}
Stolyar, B.: Apple unveils the most popular iphone apps of 2019 (Dec 2019), \url{https://mashable.com/article/apple-most-popular-iphone-apps-2019/}

\bibitem{suh2007physical}
Suh, G.E., Devadas, S.: Physical unclonable functions for device authentication and secret key generation. In: 2007 44th ACM/IEEE Design Automation Conference. pp. 9--14. IEEE (2007)

\bibitem{reqd_security}
Toponce, A.: Strong passwords need entropy (2011), \url{https://pthree.org/2011/03/07/strong-passwords-need-entropy/}

\bibitem{wiki:googleplay}
{Wikipedia contributors}: List of most-downloaded google play applications --- {Wikipedia}{,} the free encyclopedia. \url{https://en.wikipedia.org/w/index.php?title=List_of_most-downloaded_Google_Play_applications&oldid=962291709} (2020), [Online; accessed 5-July-2020]

\bibitem{CommonPasswords}
Winder, D.: Ranked: The world’s top 100 worst passwords. Available at \url{https://www.forbes.com/sites/daveywinder/2019/12/14/ranked-the-worlds-100-worst-passwords/#54064d4869b4} (2019)

\bibitem{Pswd_auth}
WordFence: Password authentication and cracking (Jun 2018), \url{https://www.wordfence.com/learn/how-passwords-work-and-cracking-passwords/}

\bibitem{yan2004password}
Yan, J., Blackwell, A., Anderson, R., Grant, A.: Password memorability and security: Empirical results. IEEE Security \& privacy  \textbf{2}(5),  25--31 (2004)

\bibitem{zetter_2017}
Zetter, K.: It's insanely easy to hack hospital equipment (Jun 2017), \url{https://www.wired.com/2014/04/hospital-equipment-vulnerable/}

\bibitem{6805770}
{Zhang-Kennedy}, L., {Chiasson}, S., {Biddle}, R.: Password advice shouldn't be boring: Visualizing password guessing attacks. In: 2013 APWG eCrime Researchers Summit. pp. 1--11 (2013)

\end{thebibliography}

\appendix

\section{Cryptographic Security}\label{sec:appendix}

\textbf{Given the limitations imposed by the very nature of algorithms optimized for humans (which are intentionally difficult to represent on a computer) these methods cannot be used directly; we use approximate, illustrative calculations to indicate the likelihood of a given scheme satisfying some property.}

When an adversary attempts to guess a user's password for random accounts after seeing $m/\lambda$ other random (account, password) pairs for the same user, a hash function $h_R$ is considered UF-RCA (Unforgeability Under Random Challenge Attack) secure if a poly-time adversary can guess a new (account, password) pair with negligible success probability. \cite{blocki2014towards} 

For any hash function $h_R(s,w_i) \longrightarrow y_i$ the adversary attempts to either guess $s$, or guess $y_j$ for some $w_j$, based on knowledge of $C = \{(y_1,w_1),(y_2,w_2),$ $\dots ,(y_n,w_n)\}$ where $(y_j,w_j) \not\in C$. The probability of correctly guessing the output (hash) for website $w_j$ without knowing $s$, i.e., $P( (y_j,w_j) | y_j = h_R(s,w_j) \land (y_j,w_j) \not\in C )\leq \epsilon$ for any probabilistic polynomial time adversary.

\subsection{Pre-image resistance}
Given only $h_R$ (public hash function) and $h_R(w,s_k)$ (a password), pre-image resistance requires that it must be computationally hard to deduce $s_k$, the subkey, and $s$, the master secret. Note that $R$ is unclonable in our setup.

\noindent
\textbf{Memory Palace}:
Given the hash, every alternate letter is either (section~\ref{sec:function-descriptions}):
\begin{itemize}
    \item $l = \mathbb{S}(x,y)$ where $\mathbb{S}$: sum and $x$, $y$ are two letters
    \item a diagonal mapping of $l$ on the keyboard
\end{itemize}
Every letter $l$ in the subkey depends on two other letters $x$, $y$ such that \footnote{assuming the alphabet is indexed from 0}:

$L 
\begin{cases}
x+y&\text{for $x+y<26$}\\
x+y-26&\text{for $x+y\geq26$}\\
\end{cases}$

The probability of guessing $x$ and $y$ given $l$ is $P(x,y | l) \leq \frac{1}{13} (0.0769) \text{ or }$ $ \frac{1}{14} (0.0714)$ based on 13-14 pairs of $s(x,y)$ for every l. This reveals nothing about the permutation, \textit{e.g.}, $a_i+b_i = b_i+a_i = c_i$, where $a_i$ is the index of the letter $a$. In this case both $ab$ and $ba$ are candidate permutations for $c$, as are 13 other letter-pairs such as $cz$, $no$ etc. So, every character of the hash depends on several possible letter-pairs in the previous text (confusion). Taking into account letter-pair permutations, the probability space increases such that: $P(x,y|l) \leq \frac{1}{26} (0.0385) \text{ or } \frac{1}{28} (0.0357)$. The adversary now guesses the underlying letters with $\leq 4\%$ probability. If all (x,y) and their permutations are discovered, the user’s subkey is discovered. However this does not reveal other subkeys due to sources of randomness within the function, as elaborated in  appendix \ref{sec:appendix}.

\noindent
\textbf{Song Password}:
Passwords generated by this method had no identifiable words from the English language, or local languages. 
The title word of the song for the examples used in Step 3 in the description of Song Password formed a maximum of 10\% of the song lyrics. An adversary has to undo several layers of confusion based on R, such as shifting characters to different positions, removing characters etc, which leave no identifiable words from the English language, or local languages in the final password, to deduce $s$ from the hash. It is also computationally hard to predict characters that may have been removed due to character shifts before deletion that do not preserve letter frequencies or word patterns.

\textbf{Scrambled Box}
This method is strongly resistant to pre-image attacks (a public S-box degrades gracefully). Given the S-box and the password, each character $c$ in the S-box corresponds to a unique coordinate set $(x,y)$ which in turn is the index $xy$ of an alphabet. If the letter maps to a single-digit index, $y$ may be a digit from the index of the next alphabet. Due to the vast number of possibilities for each character mapping in the password, we propose that finding $s$ given the user's S-box, $w$ and $h(s,w)$, is computationally infeasible.

\noindent
\textbf{Internal Sentence}: Here, $s_k=s$ is a ``unique'' word, and $h_R(s,w)$ is a sentence including $s$ and $w$. A frequency analysis of words will suggest a candidate $s$, and $w$ is publicly known. Passwords resulting from this hashing method carried high entropy, but most passwords (138/202) with 4-17 words, included between 1-12 words from the 3000 most frequently used English words \cite{press} and thus are not UF-RCA secure, as with $n$ (account, password) pairs for the same user, a ``unique'' $s$ can be deduced with word frequency analysis. Combined with word permutations a large number of candidate passwords can be produced with negligible computational effort.
However this method is still weakly collision-free - long sentences without specified one-way mappings of (subkey $\longrightarrow$ word combinations) result in a low incidence of $h_R(m')=h_R(m)$.

\subsection{Collision resistance and Randomness}

An adversary cannot even compute or verify $h_R$ efficiently, since $R$ is unique to each user. In that sense, our hash function is similar to a physically unclonable function~\cite{suh2007physical}. Our analysis suggests that given a password $y$, guessing $m$, $P(h_R(m) = y) \leq \epsilon = {2^{-78}}$ in the average case (length 11.83), as analysed at the end of section~\ref{sec:subuserstudy}. (Most of our functions are also strongly collision free; details omitted for brevity.) We refer readers to \cite{CuePinSelect} for the security of Cue-Pin-Select.

We observe a variety of sources of randomness for each $R$. Understanding and manipulating this randomness is an interesting problem for future research.

\end{document}